%% file: main_aamas.tex
\documentclass[sigconf]{aamas} 
\usepackage{balance} % for balancing columns on the final page

\setcopyright{none}
\acmConference[ALA '23]{Proc.\@ of the Adaptive and Learning Agents Workshop (ALA 2023)}
{May 9-10, 2023}{Online, \url{https://ala2023.github.io/}}{Cruz, Hayes, Wang, Yates (eds.)}
\copyrightyear{2023}
\acmYear{2023}
\acmDOI{}
\acmPrice{}
\acmISBN{}
\usepackage{microtype}
\usepackage{graphicx}
\usepackage{booktabs} % for professional tables
\usepackage{caption}
\usepackage{subcaption}
\usepackage{lipsum}
\usepackage{gensymb}

\usepackage{hyperref}
\usepackage{amsmath}
\usepackage{mathtools}
\usepackage{amsthm}

\usepackage[capitalize,noabbrev]{cleveref}

\graphicspath{{figures/}}

\acmSubmissionID{???}

\title[AAMAS-2023 Formatting Instructions]{Heterogeneous Social Value Orientation Leads to Meaningful Diversity in Sequential Social Dilemmas}

\author{Udari Madhushani}
\affiliation{
  \institution{Princeton University}
  \city{}
  \country{}}
\email{udarim@princeton.edu}
\author{Kevin R. McKee}
\affiliation{
  \institution{DeepMind}
\city{}
  \country{}}
\email{kevinrmckee@google.com}

\author{John P. Agapiou}
\affiliation{
  \institution{DeepMind}
  \city{}
  \country{}}
\email{jagapiou@google.com}

\author{Joel Z. Leibo}
\affiliation{
  \institution{DeepMind}
  \city{}
  \country{}}
\email{jzl@google.com}

\author{Richard Everett}
\affiliation{
  \institution{DeepMind}
  \city{}
  \country{}}
\email{reverett@google.com}

\author{Thomas Anthony}
\affiliation{
  \institution{DeepMind}
  \city{}
  \country{}}
\email{twa@google.com}

\author{Edward Hughes}
\affiliation{
  \institution{DeepMind}
  \city{}
  \country{}}
\email{edwardhughes@google.com}

\author{Karl Tuyls}
\affiliation{
  \institution{DeepMind}
  \city{}
  \country{}}
\email{karltuyls@google.com}

\author{Edgar A. Duéñez-Guzmán}
\affiliation{
  \institution{DeepMind}
  \city{}
  \country{}}
\email{duenez@google.com}

\begin{abstract}
\input{AAMAS_sections/1_abstract}
\end{abstract}

\begin{document}
%%% The following commands remove the headers in your paper. For final 
%%% papers, these will be inserted during the pagination process.

\pagestyle{fancy}
\fancyhead{}

%%% The next command prints the information defined in the preamble.

\maketitle 

%%%%%%%%%%%%%%%%%%%%%%%%%%%%%%%%%%%%%%%%%%%%%%%%%%%%%%%%%%%%%%%%%%%%%%%%
% Include paper content from external files 
\section{Introduction}
\input{AAMAS_sections/2_introduction}

\section{Methods}
\input{AAMAS_sections/4_method}

\section{Results}
\input{AAMAS_sections/5_results}

\section{Discussion}
\input{AAMAS_sections/7_discussion}

\pagebreak
\section*{Acknowledgement}
Authors thank Daniel Hennes, DJ Strouse, Marc Lanctot, Alexander (Sasha) Vezhnevets, Julien Perolat, Andrea Tacchetti and Ian Gemp for helpful discussion.

% Bibliography components
\bibliography{template_refs}
\bibliographystyle{ACM-Reference-Format}

\end{document}

%% file: AAMAS_sections/1_abstract.tex
In social psychology, Social Value Orientation (SVO) describes an individual's propensity to allocate resources between themself and others. In reinforcement learning, SVO has been instantiated as an intrinsic motivation that remaps an agent's rewards based on particular target distributions of group reward. Prior studies show that groups of agents endowed with heterogeneous SVO learn diverse policies in settings that resemble the incentive structure of Prisoner's dilemma. Our work extends this body of results and demonstrates that (1) heterogeneous SVO leads to meaningfully diverse policies across a range of incentive structures in sequential social dilemmas, as measured by task-specific diversity metrics; and (2) learning a best response to such policy diversity leads to better zero-shot generalization in some situations. We show that these best-response agents learn policies that are conditioned on their co-players, which we posit is the reason for improved zero-shot generalization results.

%% file: AAMAS_sections/2_introduction.tex
In psychology research, Social Value Orientation (SVO) is a cognitive construct reflecting a person's preference for resource allocation between themselves and others \cite{griesinger1973toward,liebrand1988ring,murphy2011measuring}. While some individuals may solipsistically focus on maximizing their personal success, others demonstrate different motivations, including maximizing the difference between their own and others' outcomes (a competitive orientation), maximizing collective welfare (a prosocial orientation), or maximizing other peoples' benefit (an altruistic orientation). In artificial intelligence research, various algorithms draw inspiration from these insights in their design or implementation \cite{mckee2020social,schwarting2019social}. In reinforcement learning, SVO is an intrinsic motivation that transforms an agent's reward based on a parameterized target distribution between its reward and the reward of others. Recently, studies have investigated the role of SVO in social dilemmas, situations where a group of agents or players interact in ways that involve trade-offs between their self-interest and the collective interest of the group. This research has generated insight into the impact of SVO on the emergence of diverse behaviors and cooperation \cite{mckee2020social,mckee2022quantifying}, and partner choice \cite{mckee2022warmth}. SVO research has focused primarily on social dilemmas with underlying incentive structures resembling the \emph{Prisoner's dilemma}~\cite{rapoport1974prisoner}, wherein each player has an incentive to defect, even though they would be better off if they both cooperated.

Sequential social dilemmas are a class of social dilemmas in which the decision-making process of the interacting agents is temporally and spatially extended~\cite{leibo2017multi}. Performing well in a sequential social dilemma can be accomplished by considering of long-term consequences, interdependence, and cooperation among group members. Sequential social dilemmas have been widely studied in the context of emergence and maintenance of cooperation \cite{lerer2017maintaining,peysakhovich2018consequentialist}, inequity aversion \cite{hughes2018inequity}, partner choice \cite{duenez2021statistical,mckee2022warmth}, and generalization \cite{mckee2022quantifying,agapiou2022melting} wherein agents interact with novel co-players in test scenarios.

While environments provide an \emph{extrinsic reward} that can be used to learn a policy, it is often useful to provide agents with an \emph{intrinsic reward} to shape their behavior towards a policy with desirable properties. Intrinsic reward has be used analogously to social preferences in human decision making. In most research on sequential social dilemmas, all players either have no \emph{intrinsic reward}, or they all have the same function (i.e. they have homogeneous social preferences)~\cite{lerer2017maintaining,wang2018evolving}. However, it has been observed that having a population of agents who differ in their intrinsic reward function (i.e. they have heterogeneous social preferences) can lead to higher levels of cooperation~\cite{hughes2018inequity}.
In \cite{mckee2020social, mckee2022quantifying, mckee2022warmth}, the authors showed that heterogeneity can produce behavioral diversity in group dilemmas, and in games with incentive structures similar to the Prisoner's dilemma. Other incentive structures have not yet been explored. In addition, the precise interplay between diversity in social preferences and in agent policies, particularly on the mechanisms that enable generalization to novel social partners, remains under-explored.

Diversity in policies has been demonstrated to improve various aspects of agent performance, such as exploration \cite{zahavy2022discovering}, adaptation to environmental changes \cite{derek2021adaptable}, positive group outcomes \cite{mckee2020social,tang2021discovering}, generalization to novel co-players \cite{lupu2021trajectory}, and collaboration with humans  \cite{strouse2021collaborating}. One way to quantify diversity is to examine the reward an agent obtains when interacting with different co-players (often called \emph{strategic diversity}) \cite{balduzzi2019open,garnelo2021pick}. To complement these methods, diversity can also be evaluated through state-action variation, which measures the distribution of state-action pairs that an agent traverses. State-action diversity can be assessed by measuring differences in the state visitation frequency \cite{zahavy2022discovering}, action selection frequency in a given state \cite{mckee2022quantifying}, or differences between state-action trajectories starting from a specific state \cite{lupu2021trajectory}. 
Defining an environment agnostic metric based on state-action variation that captures \emph{meaningful} diversity---that is, diversity that has a broader effect on group trajectories---can be challenging. An alternative is to instead use environment-specific measures of diversity, which the researcher can design using their knowledge of specific environment features.

Zero-shot generalization \citep{hu2020other,hu2021off,strouse2021collaborating,leibo2021scalable,mckee2022quantifying} seeks to develop general agents that are capable of successfully interacting with novel agents during test time (i.e., agents not seen during training). In such situations, the policies of the novel agents encountered at test time can be out-of-distribution for the agents, leading to poor coordination in purely cooperative settings \cite{hu2020other, lupu2021trajectory}, and getting exploited in competitive settings \cite{perez2021modelling}. In mixed-motive games, failure to generalize to novel agents can lead to deadweight loss by missing an opportunity to cooperate \cite{leibo2021scalable}. Learning a best response to partners/opponents with diverse policies has emerged as a promising approach to zero-shot generalization \citep{strouse2021collaborating}. The intuition behind this approach is that training with a set of diverse policies decreases the likelihood of encountering out-of-distribution policies at test time. Despite this promise these best response techniques have not yet been applied in a wide range of incentive structures.

In this work, we assess heterogeneous SVO in a range of incentive structures in sequential social dilemmas. We include temporally and spatially extended environments with an underlying structure that resembles several different matrix games: \emph{Prisoner's dilemma}; \emph{Chicken}, where both players have an incentive to choose a risky behavior, but where the worst outcome is if both choose the high risk; and \emph{Stag hunt} wherein players have a safe choice, and an incentive to coordinate on a high-reward strategy that carries a risk of costly miscoordination. Chicken and Stag hunt are equilibrium selection social dilemmas.

We extend the observation that heterogeneous SVO leads to diverse policies to the Chicken- and Stag hunt-like incentive structures.
We also show that this diversity, when leveraged via best response, can improve zero-shot generalization in equilibrium selection sequential social dilemmas. We found that best-response agents adapted to partners/opponents with diverse behaviors by learning a conditional policy. However, when the sequential social dilemma was not an equilibrium-selection problem, the learned best response collapsed to one unconditional policy, leading to poor zero-shot generalization

The paper is organized as follows. Section 2 outlines the methodology employed in the paper. In Subsection 2.1, we present the formulation of the $N$-agent partially observable Markov process used in the paper. Subsection 2.2 describes the Social Value Orientation (SVO) framework and its implementation. In Subsection 2.3, we discuss the various environments used in the study and their characteristics. Subsection 2.4 details the procedure for generating diverse policies in sequential social dilemmas. In Subsection 2.5, we present the process for training a best response agent with a population of agents and evaluating zero-shot generalization performance. Furthermore, we provide a description of the agent's architecture in Subsection 2.5. Section 3 presents the results of the work. In Subsection 3.1 and 3.2, we present the results obtained from generating diverse policies in environments with different incentive structures. In Subsection 3.3, we present the results of zero-shot generalization performance evaluation. Finally, in Section 4, we provide additional discussions and conclusions. The section summarizes the main contributions of the work and discuss potential societal impacts.

%% file: AAMAS_sections/4_method.tex
\begin{figure}[h!]
    \centering
     \includegraphics[width=\linewidth]{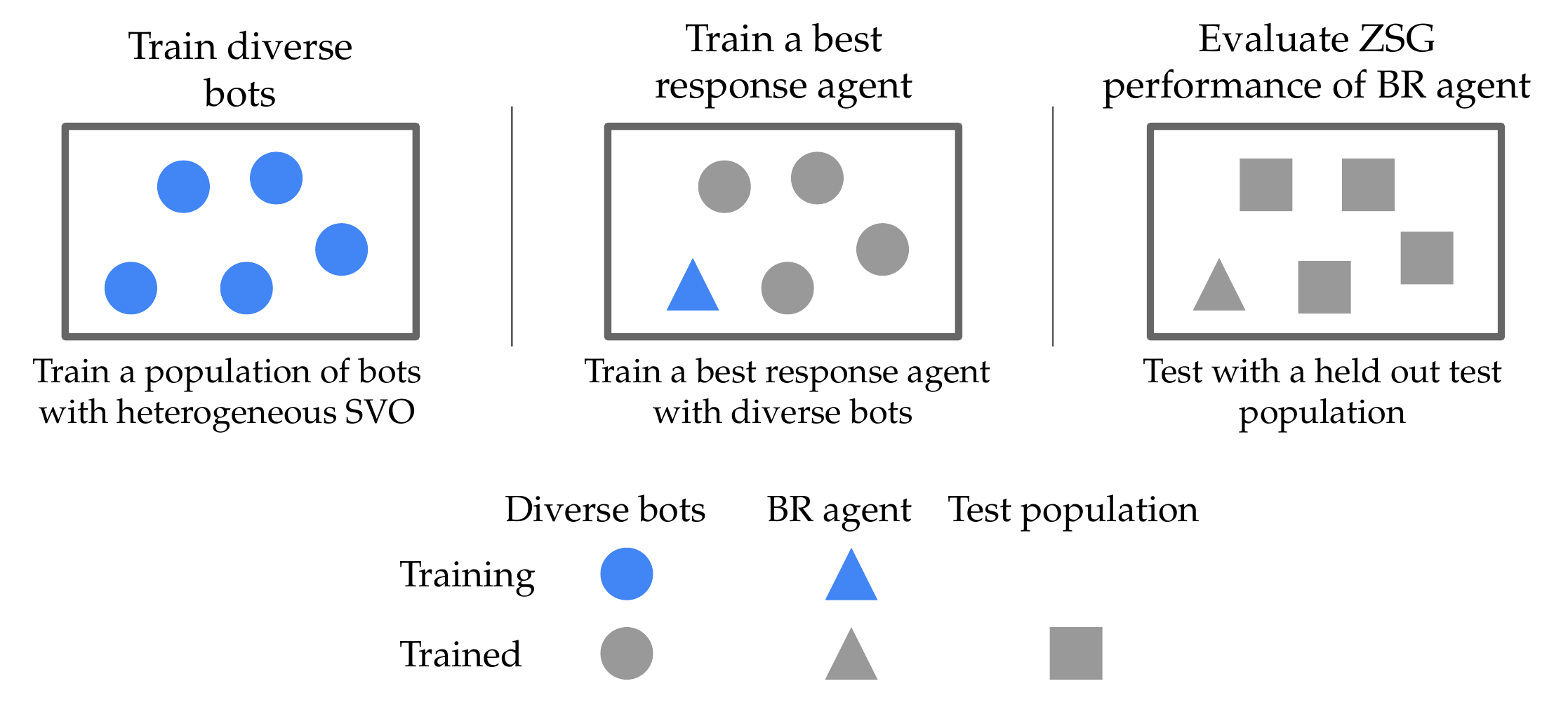}
     \caption{Overview of the methodology. Blue shapes show agents that are actively being trained, whereas gray ones denote frozen agents (bots). Circles represent the agents trained with diverse SVO, triangles denote a best response agent, and squares denote a held-out set of co-players. Evaluation is zero-shot, meaning the best response agent is frozen (gray triangle) and is evaluated against the held-out bots.}
     \label{fig:pipeline}
\end{figure}

\subsection{$N$-agent POMDP}

We consider a multi-agent partially observable Markov decision process defined by the tuple $\Big\langle N, \mathcal{S},\mathcal{A}, \mathcal{R}, P, \gamma \Big\rangle,$ where $N$ is the number of agents, $\mathcal{S}$ is the joint state space, $\mathcal{A}= \times_{i=1}^N\mathcal{A}^i$ is the joint action space, $P$ is the state transition probability distribution, $\mathcal{R}$ is the reward function and $\gamma$ is the discount factor. This can also be referred to as a partially observable Markov game \cite{littman1994markov} or a partially observable stochastic game \cite{shapley1953stochastic}. At each time step $t$, each agent $i \in {1,\ldots,N}$ observes a private (local) observation $o_t^i$ and takes an action $a_t^i$ from a set of actions $\mathcal{A}^i$. The joint action of all agents at time step $t$ is denoted as $a_t = (a_t^1, \ldots, a_t^N)$. The state $s_t$ is not observed directly by the agents, instead the partial observation $o_t^i$ depends on the current state of the environment $s_t$ and the agent's observation function. The observation function for agent $i$ is denoted as $O^i(o_t^i|s_t)$. Each agent $i$ receives a reward $r_t^i$ which is a function of the joint action $a_t$ and the state $s_t$ of the environment. The state of the environment transitions according to a probability distribution $P(s_{t+1}|s_t, a_t)$.

The objective of each agent $i$ is to maximize their cumulative expected discounted reward, over a given finite time horizon, defined as $J^i = \mathbb{E}\left[\sum_{t=0}^{T} \gamma^t r_t^i\right]$, where $\gamma \in [0,1]$ balances the importance of immediate and future rewards. The agents' policies are defined as the mapping from the agent's observation history to an action, i.e., $\pi^i(a_t^i|o_{1}^i, \cdots, o_t^i)$. The policies are updated using a multi-agent reinforcement learning algorithm that maximizes the agents' objective functions. 

\subsection{Social Value Orientation}
Omitting the dependence on $t$, let $r^i$ be the reward of agent $i.$ Let $\bar{r}^{-i}$ be the average reward of all the agent except agent $i.$ Then we have
$$
\bar{r}^{-i}=\frac{1}{N-1}\sum_{j=1, j\neq i}^Nr^j.
$$

Let $\theta^i$ denote the SVO target angle of agent $i$. Following the definition given in \cite{mckee2022warmth}, we define the effective reward $\hat{r}^i$ of agent $i$ as
$$
\hat{r}^i=r^i\cos(\theta^i) + \bar{r}^{-i}\sin(\theta^i).
$$
While sometimes intrinsic rewards are temporally smoothed (e.g.\cite{hughes2018inequity}), in this work, effective reward does not include any temporal smoothing. Reintroducing the time step $t$ from the previous section the objective function agent $i$ optimizes for is
$$
\hat{J}^i = \mathbb{E}\left[\sum_{t=0}^{T} \gamma^t \hat{r}_t^i\right].
$$

\subsection{Environments}

We provide a brief description of the environments. For all experiments in this paper, we use environments from Melting Pot 2.0 without modifications \cite{agapiou2022melting}.\\

\begin{figure}[h!]
    \centering
     \includegraphics[width=\linewidth]{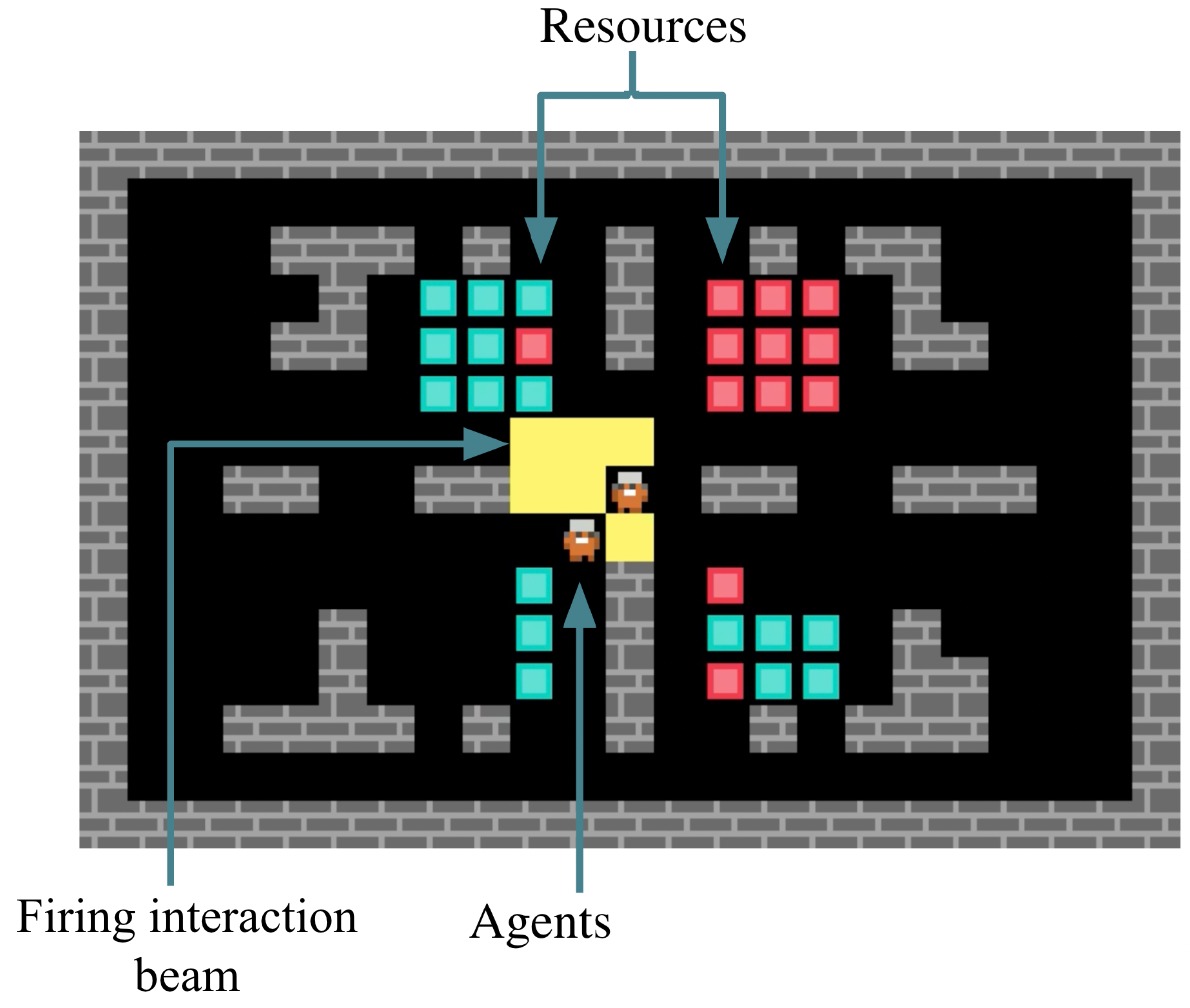}
     \caption{``In the matrix'' repeated games. This is a 2-player game where agents can gather 2 types of resources (green corresponding to cooperation, red corresponding to defection). When agents interact (using an interaction beam) they get rewards according to their inventory counts and a game specific payoff matrix. The payoff matrix can be Stag hunt, Chicken or Prisoner's dilemma type payoff matrix}
     \label{fig:ITM_env}
\end{figure}

\noindent{\bf{``in the matrix'' repeated games:}}
The ``in the matrix'' repeated games are a family of sequential social dilemmas in Melting Pot 2.0 where two-players interact. In the beginning of each episode the environment is initialized according to a given resource layout, and a set of fixed points where players can spawn. The map consists of two types of resources which can be distinguished by their colour; red corresponds to defection and green corresponds to cooperation (see Figure \ref{fig:ITM_env}). Players can pick up resources by walking over them, and these resources go into a player inventory. Players spawn with one of each resource type in their inventory. After spawning, each player can move around the map, collect resources, and interact with the co-player by firing an interaction beam. When players interact (by one player hitting the other using their interaction beam), each player gets a reward equal to the expected payoff calculated from the inventory counts and environment-specific payoff matrix. The agent who zaps the other agent is considered as the row player. The inventory count of each player defines a mixed strategy where the probability of playing each pure strategy is equivalent to the percentage of the corresponding resource. Let $N_r^i$ and $N_g^i$ denote the inventory count, number of red resources and green resources respectively, for agent $i\in {1, 2}.$  For each agent $i$ their mixed strategy is given as
$$p=\Big[\frac{N_r^i}{N_r^i+N_g^i}, \frac{N_g^i}{N_r^i+N_g^i}\Big]$$
Let $A$ be the payoff matrix for the game. Let $r_{row}$ and $r_{col}$ be the reward of row player and column player respectively. Let $p_{row}$ and $p_{col}$ be the mixed strategy probability vector of row player and column player respectively. Then the rewards can be defined as
$$r_{row}=p_{row}^TAp_{col}, \:\:\: r_{col}=p_{col}^TA^Tp_{row}$$
These reward calculations correspond to those used in game theory for matrix games and iterated social dilemmas~\cite{weibull1997evolutionary}. 

\begin{figure}[h!]
    \centering
     \includegraphics[width=\linewidth]{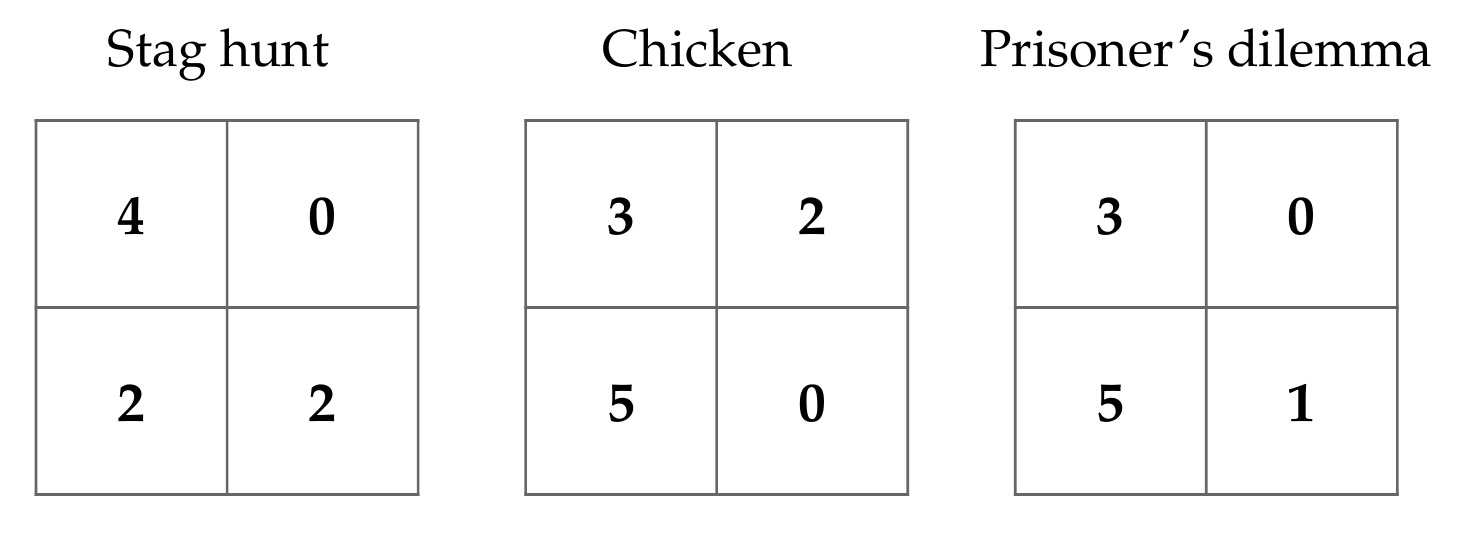}
     \caption{Payoff matrices for Stag hunt, Chicken and Prisoner's dilemma. The values shown correspond to the payoff of the row player. The payoff of the column player is the transpose of the shown matrix (i.e. the games are symmetric games). Cooperation corresponds to the first row and column. Defection corresponds to the send row and column.}
     \label{fig:payoff_matrix}
\end{figure}

The payoff matrices $A$ used are given in Figure \ref{fig:payoff_matrix}. 
After interacting, players receive their reward from interaction, freeze for 16 steps, and have their inventory counts reset (to one of each resource type). And the end of the 16 steps players disappear and get re-spawned after 5 steps. 
Players can have multiple interactions within an episode. Once a resource is picked up, it begins to regenerate after a delay of $10$ steps, with a 20\% chance of regenerating on each subsequent step. As is standard in Melting Pot 2.0, in each game, there is a 10\% chance that the episode will end after every $100$ steps, with a minimum of $1000$ steps per episode.

\noindent{\bf{Externality mushrooms:}} 
Externality mushrooms is a sequential social dilemma where players are immediately affected from prosocial or antisocial behaviors of their co-players. This is a 5-player game where players eat mushrooms in order to receive rewards. Four types of mushrooms grow (in different amounts) on the map: red, green, blue, and orange. Eating a red (``fize'': full internality zero externality) mushroom gives a reward of $1$ to the player who consumed the mushroom. Eating a green (``hihe'': half internality half externality) mushroom gives a total reward of $2/5$ \emph{to all players}. Eating a blue (``zife'': zero internality full externality) mushroom gives a total reward of $3/4$ divided equally among all players \emph{excluding the player who consumed it}. Eating an orange (``nize'': negative internality zero externality) mushroom causes red fize mushrooms to be destroyed, each with probability 0.25, and gives a reward of $-0.1$ to the player who consumed it. After eating a mushroom, the player who consumed it freezes for the mushroom's digestion time: $0$ (red), $10$ (green), $15$ (blue), and $15$ steps (orange). After spawning, a mushroom is removed from the map after its perishing time, i.e. the time it takes for the mushroom to disappear: $200$ (red), $100$ (green), and $75$ steps (blue). Orange mushrooms never perish. Mushrooms respawn from spores depending on consumption of other mushrooms. Eating a red, green, or blue mushroom releases 3 spores for red mushrooms, each spore will spawn a mushroom with probability $0.25$. Eating a green or blue mushrooms also releases $3$ spores for green mushrooms which spawn with probability $0.4$. Eating a blue mushroom also releases a blue spore which spawn with probability $0.6$. Eating an orange mushroom releases a spore for a new orange mushroom which spawns with probability $1$. Similar to ``in the matrix`` repeated games, in Externality mushrooms each episode runs for at least 1000 steps. Following that the episode terminates with probability $0.2$ at every $100$ steps.

Externality mushrooms has an incentive structure similar to Chicken, where reward is maximized selfishly by consuming red mushrooms while the others are consuming blue or green mushrooms. But if everyone else is eating red mushrooms, the selfish strategy is to eat green mushrooms, as otherwise all mushrooms would be eventually depleted.

\begin{figure}[h!]
    \centering
     \includegraphics[width=\linewidth]{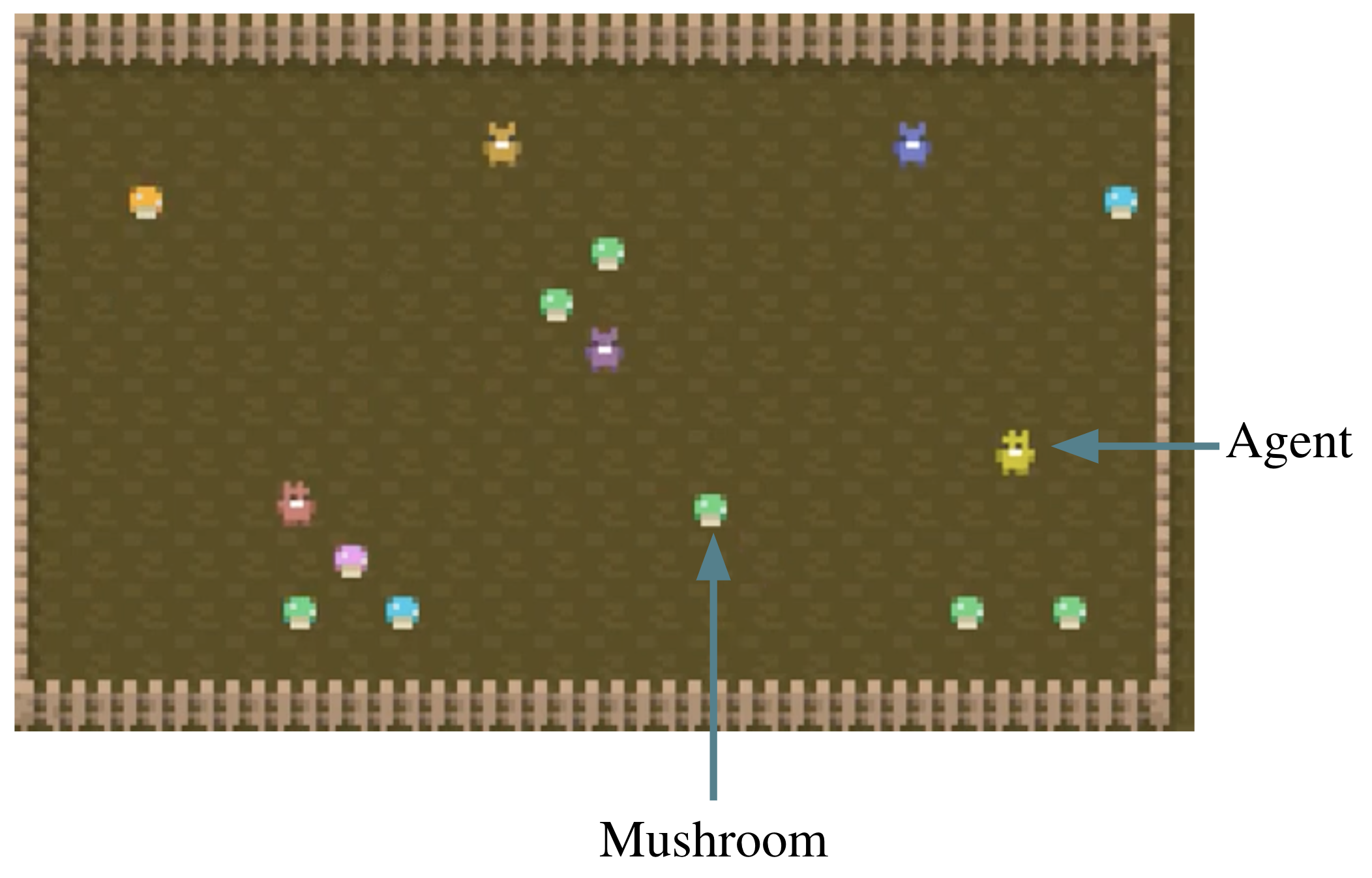}
     \caption{Externality Mushrooms. This is a 5-player sequential social dilemma game with immediate feedback. Agents instantaneously share rewards with others depending on the mushroom they are picking.}
     \label{fig:EM_env}
\end{figure}

\subsection{Generating diverse policies in sequential social dilemmas}

In the beginning of the training process we define distinct SVO angles for each agent. Each environment has a fixed number of players. We train the agents in a distributed asynchronous manner by initializing 'arenas' to train a population of agents. Arenas run in parallel and each arena is a copy of the environment with the number of players specified for that environment. This is a multi-agent version of A3C~\cite{mnih2016asynchronous} that is commonly used for multi-agent reinforcement learning~\cite{agapiou2022melting}. The Melting Pot evaluation protocol requires sampling of policies with replacement. Training in pure self-play introduces skewed reward incentives by playing with copies of oneself. To alleviate this issue, we set players in each arena to play the game for one episode either in self-play or in population-play (with equal probability). During population-play we sample agents without replacement. We train each agent for $10^9$ learner frames.

\subsection{Training a best-response agent and zero-shot generalization performance evaluation} 
We train a selfish naive learner without intrinsic reward, to best respond against the policies generated using heterogeneous SVO. In order to avoid confusion we use the term \textit{best-response agent} for the training agent, and \textit{SVO bots} for the pre-trained diverse agents trained with heterogeneous SVO values. In each episode the best-response agent plays with a set of SVO bots sampled without replacement. We train the best-response agent for $10^9$ learner frames.

Melting Pot 2.0 \citep{agapiou2022melting} provides a protocol for evaluating generalization to novel social partners, which are packaged with the suite as a held-out set of co-players in a suite of test scenarios. We measure the performance of the best-response agent using the Melting Pot test protocol. 

We use the Melting Pot test scenarios for evaluation in Stag hunt, Chicken, Prisoners' dilemma ``in the matrix `` repeated games and Externality mushrooms. Test scenario details are provided below.\\

\noindent{\bf{Test scenarios for ``in the matrix'' repeated.}}

\noindent{Focal player (our best response agent) encounters:}
\begin{itemize}
    \setlength\itemsep{5pt}
    \item [] S0: \textit{(cooperator + defector)} either a cooperator or a defector with $0.5$ probability each
    \item [] S1: \textit{(cooperator )} a cooperator
    \item [] S2: \textit{(defector)} a defector
    \item [] S3: \textit{(grim strike 1)} a player who starts by cooperating and defect for the rest the episode when focal player defects once
    \item [] S4: \textit{(grim strike 2)} a player who starts by cooperating and defect for the rest the episode when focal player defects twice
    \item [] S5: \textit{(tit-for-tat)} a player who plays tit-for-tat
    \item [] S6: \textit{(tit-for-tat tremble)} a player who a player who plays tit-for-tat and occasionally unconditionally defect. (noisy tit-for-tat)
    \item [] S7: \textit{(flipping)} a player who cooperate during the first 3 interactions and defect for the rest of the episode
    \item [] S8: \textit{(corrigible tit-for-tat)} a player who starts with defection and switch to tit-for-tat strategy when best-response agent defects
    \item [] S9: \textit{(corrigible tit-for-tat tremble)} a player who starts with defection and switch to noisy tit-for-tat strategy when best-response agent defects
\end{itemize}

\noindent{\bf{Test scenarios for Externality mushrooms:}}

\noindent{Focal player (our best response agent) encounters:}
\begin{itemize}
\setlength\itemsep{5pt}
    \item [] S0: \textit{(visiting cooperators)} 4 cooperators
    \item [] S1: \textit{(visiting defectors)} 4 defectors
\end{itemize}

\noindent{2 focal players (in our case 2 copies of best response agent) encounter:}
\begin{itemize}
\setlength\itemsep{5pt}
    \item [] S2: \textit{(resident cooperators)} 3 cooperators
    \item [] S3: \textit{(resident cooperators)} 3 defectors
\end{itemize}

We provide an overview of the end to end methodological pipeline in Figure \ref{fig:pipeline}.

\subsection{Agent architecture}

We trained the agents using the well-established Actor-Critic baseline algorithm proposed in \cite{espeholt2018impala}, building on the earlier work in \cite{mnih2016asynchronous} named Asynchronous Advantage Actor Critic or A3C.

The neural network of the agent consists of two convolutional layers, a two-layer perceptron, and an LSTM---all separated by ReLU activation functions. The convolutional layers have $16$ and $32$ output channels, kernel shapes of $8$ and $4$, and strides of $8$ and $1$. The perceptron layers are $64$ neurons each, and the LSTM layer has $128$ units. The policy and baseline for the critic are created by multilayer perceptrons ($256$ hidden units with ReLU activations) connected to the output of the LSTM.

Representation shaping is achieved through the use of an auxiliary loss and contrastive predictive coding \cite{oord2018representation}, which is used to differentiate between nearby time points via LSTM state representations. PopArt \cite{hessel2019multi} is used to adjust for the different reward scales of the different environments. The optimization method used is RMSProp with a learning rate of $4 \times 10^{-4}$, epsilon of $10^{-5}$, zero momentum, decay of $0.99$, and batch size of $256$. The baseline cost for the critic is $0.5$, and the entropy regularization cost for the policy is $0.003$.

%% file: AAMAS_sections/5_results.tex
\subsection{Experiment 1: Generating diverse policies in ``in the matrix'' repeated games}

\noindent{\bf{Experimental setup:}}
We consider Stag hunt, Chicken and Prisoners' dilemma ``in the matrix'' repeated games. For each game
we average the results over 3 random seeds. We train four agents with SVO values of $-15\degree, 0\degree, 60\degree$, and $75\degree$, respectively. These values were chosen to cluster around the incentives of competition ($-15\degree$), selfishness ($0\degree$) and pro-sociality ($60\degree, 75\degree$). The ``in the matrix'' repeated games are 2-player games. 
In addition to SVO bots we also train and freeze a set of selfish-baseline bots (i.e., no intrinsic reward) using the same procedure for comparison.

\noindent{\bf{Finding 1: Heterogeneous SVO bots learn meaningfully diverse policies}}
\vspace{5pt}
\begin{figure}[t!]
    \centering
    \includegraphics[width=\linewidth]{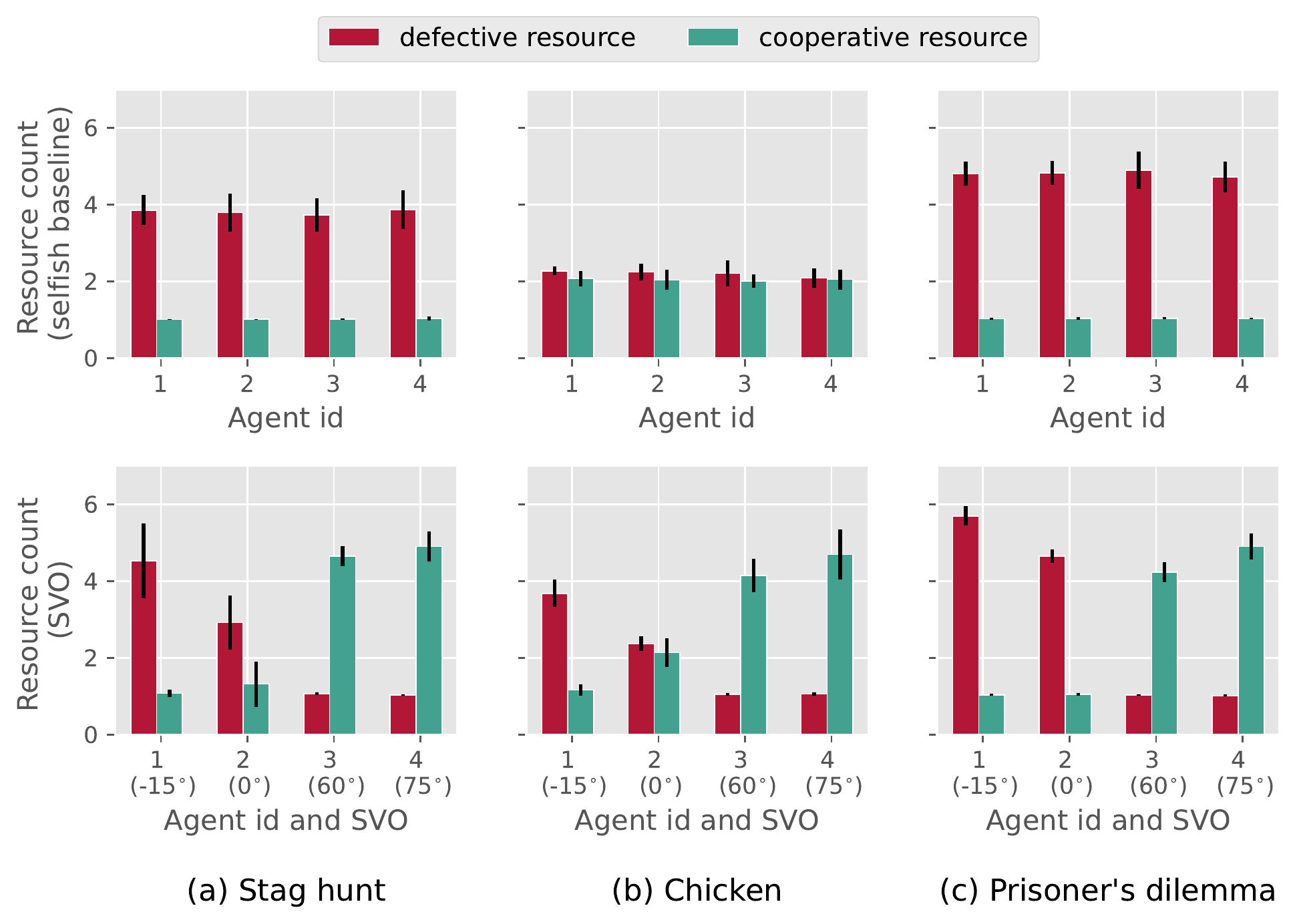}
    \caption{``In the matrix'' repeated games. {\textit{Diversity of policies of selfish-baseline bots and SVO bots.}} Each subfigure shows average inventory counts during evaluation for 4 agents, trained with 50\% self-play and 50\% population play.
    The bottom row corresponds to SVO bots with $\theta^i \in \{ -15\degree, 0\degree, 60\degree, 75\degree \}$ and the top row corresponds to selfish-baseline bots. Green and red represents cooperative and defective resource counts respectively. Error bars show the standard deviation of results over 3 random seeds.}
    \label{fig:inventory_count}
\end{figure}

We use the inventory count of the bots at the time of interaction as an environment-specific diversity measure. Since the inventory counts define the mixed strategy probability vectors, sufficiently distinct ratios of inventory counts indicate distinct mixed strategies. 
During evaluation agents play in population-play.

Figure \ref{fig:inventory_count} shows the inventory counts for the $4$ bots averaged over $500$ interactions during evaluation after the completion of training.
Top and bottom rows correspond to resource counts of selfish-baseline bots and SVO bots respectively. Figures \ref{fig:inventory_count}(a), \ref{fig:inventory_count}(b) and \ref{fig:inventory_count}(c) correspond to Stag hunt, Chicken and Prisoners' dilemma respectively. The error bars presented in the figure correspond to the average results of $3$ independent runs. The results demonstrate that in each game, all $4$ selfish-baseline bots have comparable inventory count ratios, suggesting that their policies lack diversity. Conversely, the $4$ SVO bots exhibit varied inventory count ratios, indicating diverse behaviors. For each ``in the matrix`` repeated game, resource counts correspond to SVO bots with $\theta = [-15\degree, 0\degree, 60\degree, 75\degree]$, where $\theta^i=\theta[i],$ for $i \in \{1, 2, 3, 4\}.$
We denote the cooperative resource counts and defective resource counts using green and red respectively. As the SVO angles increase from $-15\degree$ to $75\degree$, the ratio between the red and green resource counts increases, indicating more altruistic behavior.

\subsection{Experiment 2: Generating diverse policies in Externality Mushrooms}

\noindent{\bf{Experimental setup:}}
Similar to the training process in ``in the matrix`` repeated game we average the results from 3 random seeds. For each seed we train $5$ agents with SVO values of $-15\degree, 0\degree, 60\degree, 75\degree$, and $90\degree$, respectively in $50\%$ self-play and $50\%$ population-play. In addition to SVO bots we also train a set of selfish-baseline bots, using the same procedure for comparison. 

\noindent{\bf{Finding 2: The results extends to multi-player games with more than $2$ players}}
\vspace{5pt}
\begin{figure}[t!]
    \centering
     \includegraphics[width=0.8\linewidth]{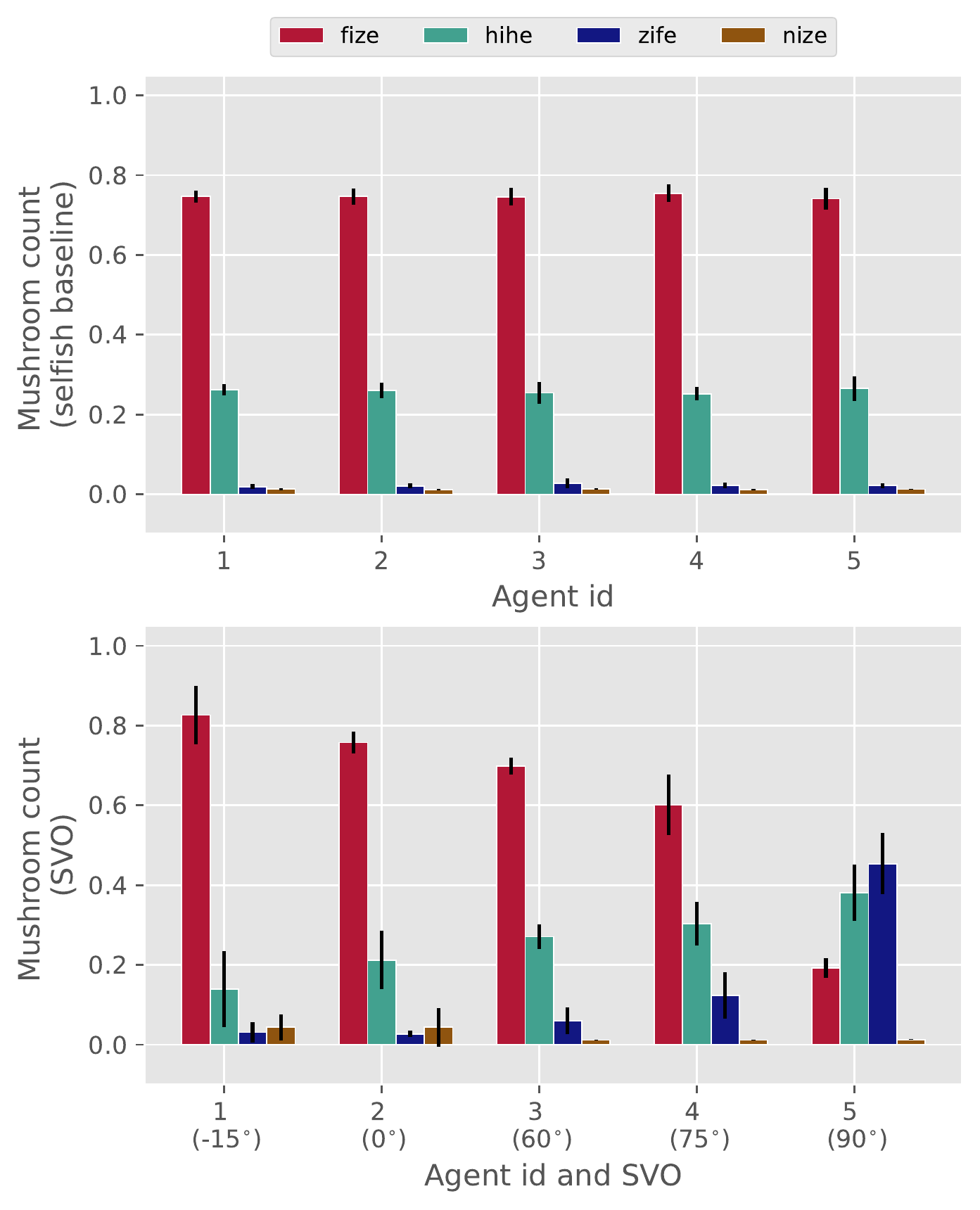}
     \caption{Externality mushrooms. {\emph{Diversity of policies of selfish-baseline bots and SVO bots.}} Each plot shows average fraction of mushrooms consumed by 5 agents during evaluation, trained with 50\% self-play and 50\% population play
     in Externality mushrooms dense game. 
     The bottom row corresponds to SVO agents with $\theta^i \in \{ -15\degree, 0\degree, 60\degree, 75\degree, 90\degree \}$ and the top row corresponds to selfish-baseline agents. Error bars show the standard deviation of results over 3 random seeds.}
     \label{fig:mushroom_count}
\end{figure}

We show that our method scales to games with more than 2 players. Figure \ref{fig:mushroom_count} shows that in Externality Mushrooms, agents trained using heterogeneous SVO learn diverse policies. We use the count of mushrooms consumed of each type as the environment-specific diversity metric. The selfish-baseline bots tend to consume mushrooms at similar ratios across different types, whereas the SVO bots consume varying ratios of different mushroom types exhibiting meaningfully diverse behaviors. Agents with low (or negative) SVO consume the selfish mushroom (red), and even the spiteful mushroom (orange), whereas those with high SVO, tend to consume more of the prosocial mushrooms (green and blue).

\subsection{Experiment 3: Zero-shot generalization evaluation}
We evaluate the zero-shot generalization performance of a learned best response to the SVO bots trained using heterogeneous SVO.
% \vspace{5pt}
\noindent{\bf{Baselines:}}
We compare the performance of a learned best response policy for SVO bots with a best response to selfish-baseline bots, Fictitious co-play (FCP, a type of best response that includes also earlier checkpoints of the agents to best respond to) \citep{strouse2021collaborating}, and exploiters (i.e., a best response agent trained on the test scenario directly) \citep{agapiou2022melting}. We train one exploiter for each test scenario. To train FCP agents we train a naive learning agent with $3$ checkpoints for each bot from a bot population. Here we use the first checkpoint, mid checkpoint and last checkpoint. The mid checkpoint is the time during training where the agent first obtains half of its final reward, of the policies of the bots. We report results for FCP applied to the heterogeneous SVO bots FCP(SVO), as well as to selfish baselines FCP(selfish-baseline). To evaluate zero-shot generalization, we also compare the performance of best response agents with the performance of selfish-baseline agents and random agents.\\  

\noindent{\bf{Experimental setup:}}
We train best-response agents for the selfish-baseline bots and SVO bots. Recall that we trained each type of bots, i.e., selfish-baseline or SVO, for 3 random seeds in this setup. We train a best-response agent for bots from each seed. For each type of test bots we show the average performance evaluation runs correspond to these $3$ training runs.

\begin{figure}[ht!]
    \centering
    \includegraphics[width=0.8\linewidth]{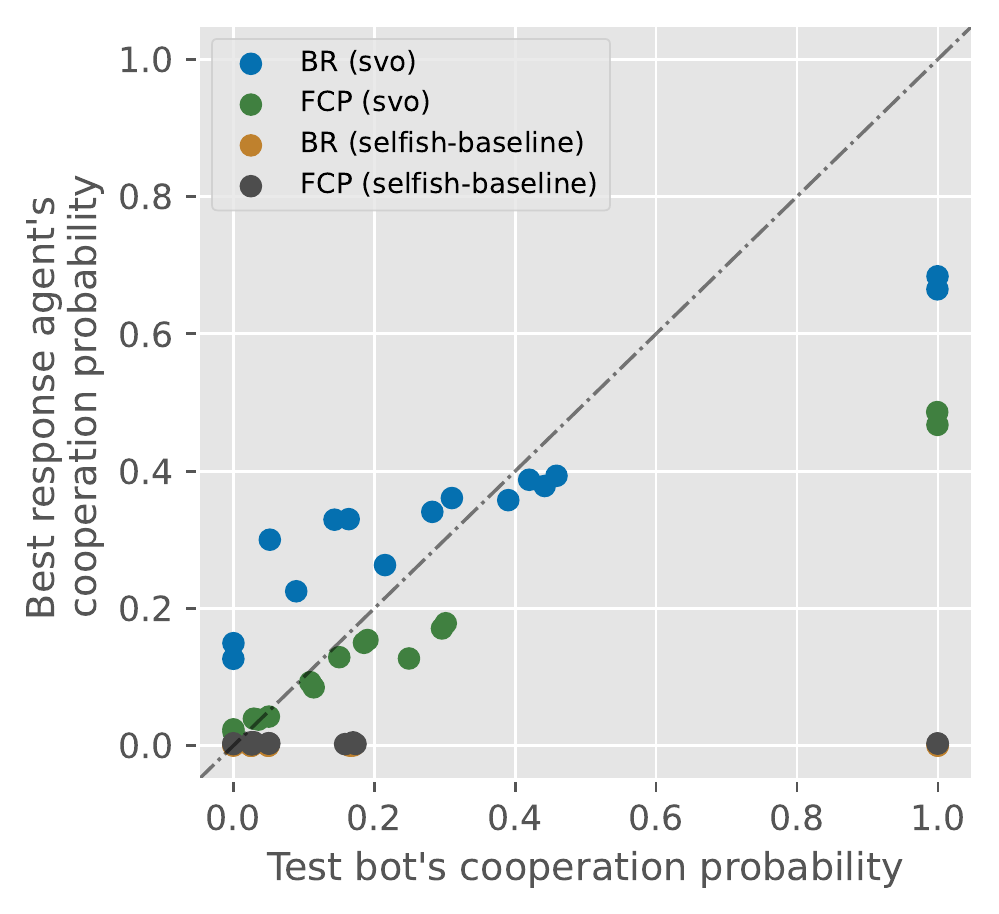}
    \caption{\textit{Comparing how well best-response agents learn conditional policies in Stag hunt in the matrix.}}
    \label{fig:ZSG_SH}
\end{figure}

\begin{figure}[htb!]
    \centering
    \includegraphics[width=0.8\linewidth]{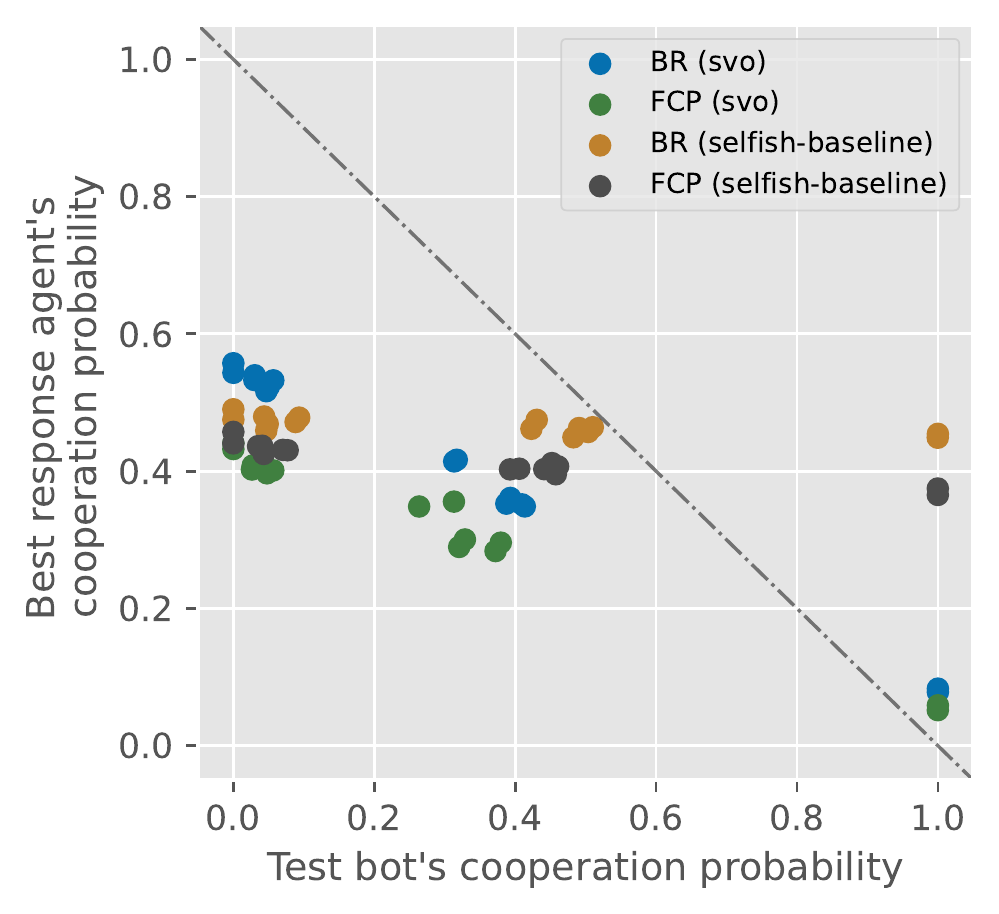}
    \caption{\textit{Comparing how well best-response agents learn conditional policies in Chicken in the matrix.}}
    \label{fig:ZSG_CH}
\end{figure}

\noindent{\bf{Finding 3: Best-response agents learn a conditional behaviour}}
In order to get a better understanding about the learned policies of the best-response agents we analyze the behaviour of the best-response agents during test time. For each test bot, Figures \ref{fig:ZSG_SH} and \ref{fig:ZSG_CH} show the fraction of interactions where the best-response agent cooperated with a bot with respect to the fraction of interactions where the bot cooperated with the best-response agent. Figure \ref{fig:ZSG_SH} corresponds to Stag hunt ``in the matrix`` repeated and \ref{fig:ZSG_CH} corresponds to Chicken ``in the matrix'' repeated. 

In this analysis we define the best-response agent's interaction as a cooperation when they have higher number of cooperative resources than defective resources in their inventory at the time of interaction. In Stag hunt in the matrix, both agents cooperating, i.e., both agents playing Stag, yields a higher reward, but it is a riskier strategy. Defecting, yields a secure payoff. Both agents cooperating or both defecting are Nash equilibria, that is, there is no incentive to unilaterally deviate from that strategy. An agent who cooperates with a defector gets 0 reward. When trained in Stag hunt in the matrix, selfish-baseline bots learn to defect. The best response to unconditional defectors is defecting. Hence the best-response agents trained with selfish-baseline bots learn to unconditionally defect. In contrast the heterogeneous SVO bot population consists of both defectors and cooperators with different levels of cooperation and defection. Best-response agents training with SVO bots encounter both cooperators and defectors and subsequently learn a conditional policy that tends to cooperate with cooperators and defect with defectors.

In Chicken in the matrix, the two Nash equilibria are for one agent to cooperate (swerve) and the other agent to defect (straight). In this case selfish-baseline agents learn to do both defection and cooperation. Hence the best-response agents trained with selfish-baseline bots also learn to defect and cooperate. However in Figure \ref{fig:ZSG_CH} we see that this behaviour is not conditional. In contrast best-response agents training with SVO bots encounter mostly cooperative and mostly defective bots, leading to best-response agents learning a conditional behavior where they tend to cooperate with defectors and defect against cooperators.\\

\begin{figure}[htb!]
    \centering
    \includegraphics[width=0.8\linewidth]{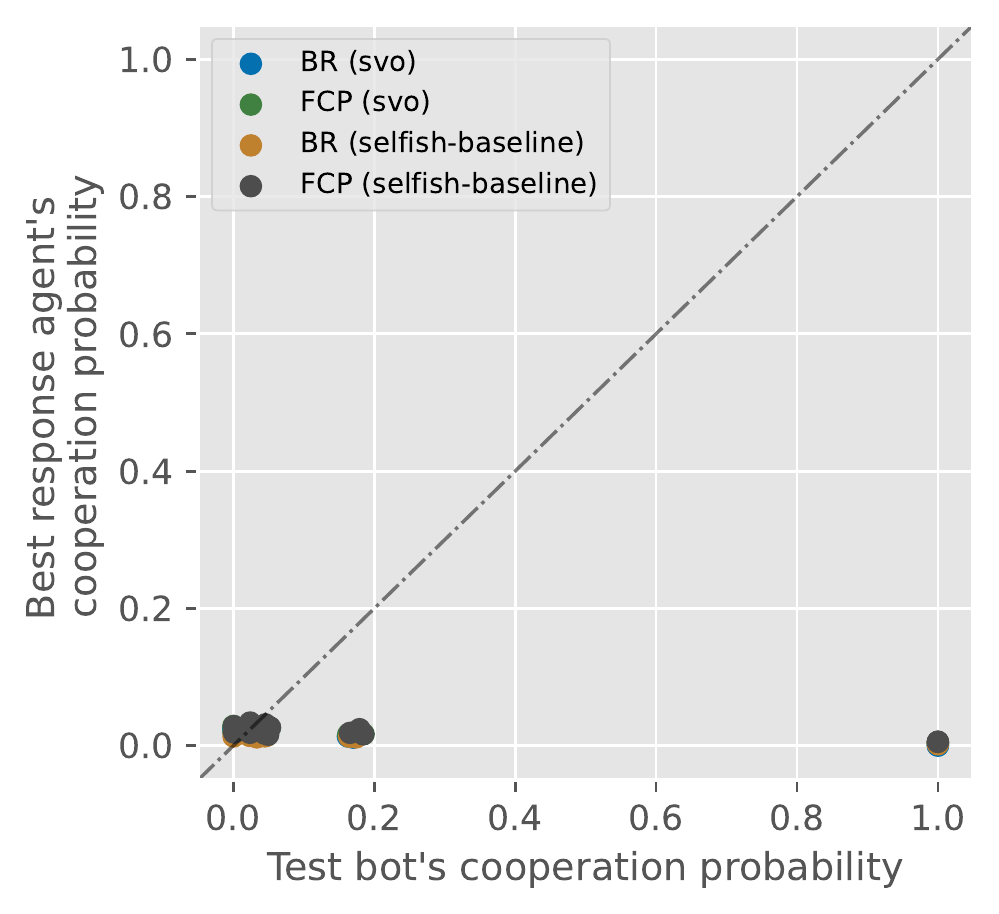}
    \caption{\textit{Comparing how well best-response agents learn conditional policies in Prisoner's dilemma in the matrix.}}
    \label{fig:ZSG_PD}
\end{figure}

\noindent{\bf{Finding 4: Failure case with Prisoner's dilemma}}
In Prisoner's dilemma in the matrix, the Nash equilibrium is both agents defecting, as a result selfish-baseline agents learn to unconditionally defect. Thus, the best response agents that are trained with selfish agents also learn to defect. Moreover, defection is also a best response to unconditional cooperation. Because SVO bots learned only unconditional strategies (either cooperate or defect), the best response to SVO bots is also to unconditionally defect.
Figure \ref{fig:ZSG_PD} illustrates this showing that all the best-response agents are learning to defect regardless of the level of cooperation of their partners. \\

\begin{table*}
  \centering
  \resizebox{0.9\linewidth}{!}{
  \begin{tabular}{@{}lccccccc@{}}
    \toprule
      & BR(SVO) & FCP(SVO) & BR(selfish-baseline) & FCP(selfish-baseline) & selfish-baseline &  random & exploiter\\ \midrule
    Stag hunt ITMR & \bf{0.876} & 0.830 & 0.856 & 0.847 & 0.850 & 0.000 & \bf{0.988} \\
    Chicken ITMR & 0.696 & 0.668 & \bf{0.745} & 0.723 & 0.723 & 0.000 & \bf{0.958} \\
    Prisoner's dilemma ITMR & 0.738 & 0.702 & 0.777 & \bf{0.783} & 0.754 & 0.000 & \bf{1.000} \\
    Externality mushrooms & 0.619 & 0.764 & 0.612 & \bf{0.846} & 0.660 & 0.000 & \bf{0.900} \\ \bottomrule
    \end{tabular}
    }
  \caption{Zero-shot generalization performance of best response agents, selfish-baseline agent, random agent and exploiter. The score is calculated by first re scaling the rewards received by each agent such that in each scenario the agent with highest(lowest) reward gets score 1(0) and then averaging over all scenarios for each environment.}
  \label{tab:ZSGtable}
\end{table*}

\noindent{\bf{Finding 5: Best response agents perform better in zero-shot generalization}} Zero-shot generalization performance of the best response agents, selfish-baseline agent, random agent, and exploiters are given in Table \ref{tab:ZSGtable}. Each agent type is run on each test scenario and their average returns are calculated. The score is normalized across agents for each scenario where the best agent receives a score of $1$, and the worst a score of $0$. The final score of an agent is their average over all scenarios. This is the same method used in Melting Pot~\cite{agapiou2022melting}. The exploiters and random agent are intended to provide approximate upper and lower bounds for performance across all environments. As expected the table shows that the exploiters achieve the best performance, while the random agent performs the worst. Across all environments at least one best response agent performs better than the selfish-baseline agent indicating that learning a best response improves zero-shot generalization.

On average in the Stag hunt in the matrix scenarios, BR(SVO) outperforms other agents. From Figure \ref{fig:ZSG_SH} we see that BR(SVO) and FCP(SVO) cooperate with unconditional defectors with a small probability. However, in Stag hunt in the matrix, an agent cooperating with a defector or defecting with a defector receives the same reward. Thus when encountering defectors and test bots that are more likely to defect BR(SVO), FCP(SVO) receives comparable rewards to BR(selfish-baseline) and FCP(selfish-baseline). When encountering more cooperative test bots, best response agents that are able to adapt to partner behaviours and cooperate with cooperators receive a higher reward. This leads to the higher score of the BR(SVO) agent in Stag hunt in the matrix.

Table \ref{tab:ZSGtable} shows that in Chicken in the matrix scenarios, BR(selfish-baseline) outperforms other agents. Note that in Chicken in the matrix, an agent cooperating with a defector receives a higher reward than an agent defecting against a defector. From results in Figure \ref{fig:ZSG_CH} we see that when test bots defect with a probability close to $1$ all the best response agents cooperate with similar probabilities. Thus in scenarios where test bots are unconditionally defecting all the best response agents obtain comparable performance. However, when test bots are cooperating with about 0.4 probability, BR(selfish-baseline) and FCP(selfish-baseline) cooperate with a higher matching probability compared to BR(SVO) and FCP(SVO) thus leading to better performance for BR(selfish-baseline) and FCP(selfish-baseline). In scenarios where best response agents encounter unconditional cooperators BR(SVO) and FCP(SVO) defect with a probability close to $1$ obtaining better performance compared to BR(selfish-baseline) and FCP(selfish-baseline). Since most of the test scenarios consist of defectors or test bots that are more likely to defect, this leads to BR(selfish-baseline) outperforming BR(SVO) and BR(FCP) agents.

Recall that Figure \ref{fig:ZSG_PD} illustrates that all the best-response agents are defecting against all test bots. Thus we expect the performance score of best response agents for Prisoner's dilemma in the matrix given in Table \ref{tab:ZSGtable} to be similar. However, surprisingly BR(selfish-baseline) and FCP(selfish-baseline) perform better than BR(SVO) and FCP(SVO). We leave investigating this as future work.

In Externality mushrooms, FCP type best response agents perform better than best response agents trained with only final policies of the co-players. This indicates that best response agents that encounter less proficient agents as well as more proficient agents perform better than the best response agents that only encounter proficient agents during training time.

%% file: AAMAS_sections/7_discussion.tex
In this paper we investigated the impact of heterogeneous Social Value Orientation on different incentive structures in sequential social dilemmas. We tested whether the presence of heterogeneous SVO leads to diverse policies and if learning a best response to these policies improves zero-shot generalization. The study found that the presence of heterogeneous SVO does indeed lead to measurable diversity in policies, and this diversity sometimes results in better zero-shot generalization for agents that best respond to them.

The best-response agents achieve better performance by learning a conditional policy that adapts to novel agents during test time. The study also revealed that when the sequential social dilemma is not an equilibrium-selection problem, this method still generates meaningful diversity in policies, but it fails to achieve better zero-shot generalization performance. This occurs because the best response to a diverse set of policies collapses to one unconditional policy that performs poorly when encountering conditional policies during test time.

Our findings have implications for understanding how heterogeneous SVO impacts incentive structures and policy diversity, and how agents can learn to adapt to diverse policies during test time to achieve better zero-shot generalization performance. Our findings provide new insights into the behavior of agents in sequential social dilemmas and highlights the importance of considering the role of heterogeneity in SVO in the design of incentive structures.

We observed that SVO agents were able to learn cooperative policies in all of the environments we tested. This hints at the potential value of using SVO to capture at least some of the aspects necessary to align agents with human values.

%% file: main_aamas.bbl

\begin{thebibliography}{33}

%%% the Web address.

\ifx \showCODEN    \undefined \def \showCODEN     #1{\unskip}     \fi
\ifx \showDOI      \undefined \def \showDOI       #1{#1}\fi
\ifx \showISBNx    \undefined \def \showISBNx     #1{\unskip}     \fi
\ifx \showISBNxiii \undefined \def \showISBNxiii  #1{\unskip}     \fi
\ifx \showISSN     \undefined \def \showISSN      #1{\unskip}     \fi
\ifx \showLCCN     \undefined \def \showLCCN      #1{\unskip}     \fi
\ifx \shownote     \undefined \def \shownote      #1{#1}          \fi
\ifx \showarticletitle \undefined \def \showarticletitle #1{#1}   \fi
\ifx \showURL      \undefined \def \showURL       {\relax}        \fi
% invisible to TeX
\providecommand\bibfield[2]{#2}
\providecommand\bibinfo[2]{#2}
\providecommand\natexlab[1]{#1}
\providecommand\showeprint[2][]{arXiv:#2}

\bibitem[\protect\citeauthoryear{Agapiou, Vezhnevets,
  Du{\'e}{\~n}ez-Guzm{\'a}n, Matyas, Mao, Sunehag, K{\"o}ster, Madhushani,
  Kopparapu, Comanescu, et~al\mbox{.}}{Agapiou et~al\mbox{.}}{2022}]%
        {agapiou2022melting}
\bibfield{author}{\bibinfo{person}{John~P Agapiou},
  \bibinfo{person}{Alexander~Sasha Vezhnevets}, \bibinfo{person}{Edgar~A
  Du{\'e}{\~n}ez-Guzm{\'a}n}, \bibinfo{person}{Jayd Matyas},
  \bibinfo{person}{Yiran Mao}, \bibinfo{person}{Peter Sunehag},
  \bibinfo{person}{Raphael K{\"o}ster}, \bibinfo{person}{Udari Madhushani},
  \bibinfo{person}{Kavya Kopparapu}, \bibinfo{person}{Ramona Comanescu},
  {et~al\mbox{.}}} \bibinfo{year}{2022}\natexlab{}.
\newblock \showarticletitle{Melting Pot 2.0}.
\newblock \bibinfo{journal}{\emph{arXiv preprint arXiv:2211.13746}}
  (\bibinfo{year}{2022}).
\newblock


\bibitem[\protect\citeauthoryear{Balduzzi, Garnelo, Bachrach, Czarnecki,
  Perolat, Jaderberg, and Graepel}{Balduzzi et~al\mbox{.}}{2019}]%
        {balduzzi2019open}
\bibfield{author}{\bibinfo{person}{David Balduzzi}, \bibinfo{person}{Marta
  Garnelo}, \bibinfo{person}{Yoram Bachrach}, \bibinfo{person}{Wojciech
  Czarnecki}, \bibinfo{person}{Julien Perolat}, \bibinfo{person}{Max
  Jaderberg}, {and} \bibinfo{person}{Thore Graepel}.}
  \bibinfo{year}{2019}\natexlab{}.
\newblock \showarticletitle{Open-ended learning in symmetric zero-sum games}.
  In \bibinfo{booktitle}{\emph{International Conference on Machine Learning}}.
  PMLR, \bibinfo{pages}{434--443}.
\newblock


\bibitem[\protect\citeauthoryear{Derek and Isola}{Derek and Isola}{2021}]%
        {derek2021adaptable}
\bibfield{author}{\bibinfo{person}{Kenneth Derek} {and}
  \bibinfo{person}{Phillip Isola}.} \bibinfo{year}{2021}\natexlab{}.
\newblock \showarticletitle{Adaptable agent populations via a generative model
  of policies}.
\newblock \bibinfo{journal}{\emph{Advances in Neural Information Processing
  Systems}}  \bibinfo{volume}{34} (\bibinfo{year}{2021}),
  \bibinfo{pages}{3902--3913}.
\newblock


\bibitem[\protect\citeauthoryear{Du{\'e}{\~n}ez-Guzm{\'a}n, McKee, Mao, Coppin,
  Chiappa, Vezhnevets, Bakker, Bachrach, Sadedin, Isaac,
  et~al\mbox{.}}{Du{\'e}{\~n}ez-Guzm{\'a}n et~al\mbox{.}}{2021}]%
        {duenez2021statistical}
\bibfield{author}{\bibinfo{person}{Edgar~A Du{\'e}{\~n}ez-Guzm{\'a}n},
  \bibinfo{person}{Kevin~R McKee}, \bibinfo{person}{Yiran Mao},
  \bibinfo{person}{Ben Coppin}, \bibinfo{person}{Silvia Chiappa},
  \bibinfo{person}{Alexander~Sasha Vezhnevets}, \bibinfo{person}{Michiel~A
  Bakker}, \bibinfo{person}{Yoram Bachrach}, \bibinfo{person}{Suzanne Sadedin},
  \bibinfo{person}{William Isaac}, {et~al\mbox{.}}}
  \bibinfo{year}{2021}\natexlab{}.
\newblock \showarticletitle{Statistical discrimination in learning agents}.
\newblock \bibinfo{journal}{\emph{arXiv preprint arXiv:2110.11404}}
  (\bibinfo{year}{2021}).
\newblock


\bibitem[\protect\citeauthoryear{Espeholt, Soyer, Munos, Simonyan, Mnih, Ward,
  Doron, Firoiu, Harley, Dunning, et~al\mbox{.}}{Espeholt
  et~al\mbox{.}}{2018}]%
        {espeholt2018impala}
\bibfield{author}{\bibinfo{person}{Lasse Espeholt}, \bibinfo{person}{Hubert
  Soyer}, \bibinfo{person}{Remi Munos}, \bibinfo{person}{Karen Simonyan},
  \bibinfo{person}{Vlad Mnih}, \bibinfo{person}{Tom Ward},
  \bibinfo{person}{Yotam Doron}, \bibinfo{person}{Vlad Firoiu},
  \bibinfo{person}{Tim Harley}, \bibinfo{person}{Iain Dunning},
  {et~al\mbox{.}}} \bibinfo{year}{2018}\natexlab{}.
\newblock \showarticletitle{Impala: Scalable distributed deep-rl with
  importance weighted actor-learner architectures}. In
  \bibinfo{booktitle}{\emph{International conference on machine learning}}.
  PMLR, \bibinfo{pages}{1407--1416}.
\newblock


\bibitem[\protect\citeauthoryear{Garnelo, Czarnecki, Liu, Tirumala, Oh, Gidel,
  van Hasselt, and Balduzzi}{Garnelo et~al\mbox{.}}{2021}]%
        {garnelo2021pick}
\bibfield{author}{\bibinfo{person}{Marta Garnelo},
  \bibinfo{person}{Wojciech~Marian Czarnecki}, \bibinfo{person}{Siqi Liu},
  \bibinfo{person}{Dhruva Tirumala}, \bibinfo{person}{Junhyuk Oh},
  \bibinfo{person}{Gauthier Gidel}, \bibinfo{person}{Hado van Hasselt}, {and}
  \bibinfo{person}{David Balduzzi}.} \bibinfo{year}{2021}\natexlab{}.
\newblock \showarticletitle{Pick Your Battles: Interaction Graphs as
  Population-Level Objectives for Strategic Diversity}. In
  \bibinfo{booktitle}{\emph{Proceedings of the 20th International Conference on
  Autonomous Agents and MultiAgent Systems}}. \bibinfo{pages}{1501--1503}.
\newblock


\bibitem[\protect\citeauthoryear{Griesinger and Livingston~Jr}{Griesinger and
  Livingston~Jr}{1973}]%
        {griesinger1973toward}
\bibfield{author}{\bibinfo{person}{Donald~W Griesinger} {and}
  \bibinfo{person}{James~W Livingston~Jr}.} \bibinfo{year}{1973}\natexlab{}.
\newblock \showarticletitle{Toward a model of interpersonal motivation in
  experimental games}.
\newblock \bibinfo{journal}{\emph{Behavioral science}} \bibinfo{volume}{18},
  \bibinfo{number}{3} (\bibinfo{year}{1973}), \bibinfo{pages}{173--188}.
\newblock


\bibitem[\protect\citeauthoryear{Hessel, Soyer, Espeholt, Czarnecki, Schmitt,
  and van Hasselt}{Hessel et~al\mbox{.}}{2019}]%
        {hessel2019multi}
\bibfield{author}{\bibinfo{person}{Matteo Hessel}, \bibinfo{person}{Hubert
  Soyer}, \bibinfo{person}{Lasse Espeholt}, \bibinfo{person}{Wojciech
  Czarnecki}, \bibinfo{person}{Simon Schmitt}, {and} \bibinfo{person}{Hado van
  Hasselt}.} \bibinfo{year}{2019}\natexlab{}.
\newblock \showarticletitle{Multi-task deep reinforcement learning with
  popart}. In \bibinfo{booktitle}{\emph{Proceedings of the AAAI Conference on
  Artificial Intelligence}}, Vol.~\bibinfo{volume}{33}.
  \bibinfo{pages}{3796--3803}.
\newblock


\bibitem[\protect\citeauthoryear{Hu, Lerer, Cui, Pineda, Brown, and
  Foerster}{Hu et~al\mbox{.}}{2021}]%
        {hu2021off}
\bibfield{author}{\bibinfo{person}{Hengyuan Hu}, \bibinfo{person}{Adam Lerer},
  \bibinfo{person}{Brandon Cui}, \bibinfo{person}{Luis Pineda},
  \bibinfo{person}{Noam Brown}, {and} \bibinfo{person}{Jakob Foerster}.}
  \bibinfo{year}{2021}\natexlab{}.
\newblock \showarticletitle{Off-belief learning}. In
  \bibinfo{booktitle}{\emph{International Conference on Machine Learning}}.
  PMLR, \bibinfo{pages}{4369--4379}.
\newblock


\bibitem[\protect\citeauthoryear{Hu, Lerer, Peysakhovich, and Foerster}{Hu
  et~al\mbox{.}}{2020}]%
        {hu2020other}
\bibfield{author}{\bibinfo{person}{Hengyuan Hu}, \bibinfo{person}{Adam Lerer},
  \bibinfo{person}{Alex Peysakhovich}, {and} \bibinfo{person}{Jakob Foerster}.}
  \bibinfo{year}{2020}\natexlab{}.
\newblock \showarticletitle{“Other-Play” for Zero-Shot Coordination}. In
  \bibinfo{booktitle}{\emph{International Conference on Machine Learning}}.
  PMLR, \bibinfo{pages}{4399--4410}.
\newblock


\bibitem[\protect\citeauthoryear{Hughes, Leibo, Phillips, Tuyls,
  Due{\~n}ez-Guzman, Garc{\'\i}a~Casta{\~n}eda, Dunning, Zhu, McKee, Koster,
  et~al\mbox{.}}{Hughes et~al\mbox{.}}{2018}]%
        {hughes2018inequity}
\bibfield{author}{\bibinfo{person}{Edward Hughes}, \bibinfo{person}{Joel~Z
  Leibo}, \bibinfo{person}{Matthew Phillips}, \bibinfo{person}{Karl Tuyls},
  \bibinfo{person}{Edgar Due{\~n}ez-Guzman}, \bibinfo{person}{Antonio
  Garc{\'\i}a~Casta{\~n}eda}, \bibinfo{person}{Iain Dunning},
  \bibinfo{person}{Tina Zhu}, \bibinfo{person}{Kevin McKee},
  \bibinfo{person}{Raphael Koster}, {et~al\mbox{.}}}
  \bibinfo{year}{2018}\natexlab{}.
\newblock \showarticletitle{Inequity aversion improves cooperation in
  intertemporal social dilemmas}.
\newblock \bibinfo{journal}{\emph{Advances in neural information processing
  systems}}  \bibinfo{volume}{31} (\bibinfo{year}{2018}).
\newblock


\bibitem[\protect\citeauthoryear{Leibo, Due{\~n}ez-Guzman, Vezhnevets, Agapiou,
  Sunehag, Koster, Matyas, Beattie, Mordatch, and Graepel}{Leibo
  et~al\mbox{.}}{2021}]%
        {leibo2021scalable}
\bibfield{author}{\bibinfo{person}{Joel~Z Leibo}, \bibinfo{person}{Edgar~A
  Due{\~n}ez-Guzman}, \bibinfo{person}{Alexander Vezhnevets},
  \bibinfo{person}{John~P Agapiou}, \bibinfo{person}{Peter Sunehag},
  \bibinfo{person}{Raphael Koster}, \bibinfo{person}{Jayd Matyas},
  \bibinfo{person}{Charlie Beattie}, \bibinfo{person}{Igor Mordatch}, {and}
  \bibinfo{person}{Thore Graepel}.} \bibinfo{year}{2021}\natexlab{}.
\newblock \showarticletitle{Scalable evaluation of multi-agent reinforcement
  learning with melting pot}. In \bibinfo{booktitle}{\emph{International
  Conference on Machine Learning}}. PMLR, \bibinfo{pages}{6187--6199}.
\newblock


\bibitem[\protect\citeauthoryear{Leibo, Zambaldi, Lanctot, Marecki, and
  Graepel}{Leibo et~al\mbox{.}}{2017}]%
        {leibo2017multi}
\bibfield{author}{\bibinfo{person}{Joel~Z Leibo}, \bibinfo{person}{Vinicius
  Zambaldi}, \bibinfo{person}{Marc Lanctot}, \bibinfo{person}{Janusz Marecki},
  {and} \bibinfo{person}{Thore Graepel}.} \bibinfo{year}{2017}\natexlab{}.
\newblock \showarticletitle{Multi-agent Reinforcement Learning in Sequential
  Social Dilemmas}. In \bibinfo{booktitle}{\emph{Proceedings of the 16th
  Conference on Autonomous Agents and MultiAgent Systems}}.
  \bibinfo{pages}{464--473}.
\newblock


\bibitem[\protect\citeauthoryear{Lerer and Peysakhovich}{Lerer and
  Peysakhovich}{2017}]%
        {lerer2017maintaining}
\bibfield{author}{\bibinfo{person}{Adam Lerer} {and} \bibinfo{person}{Alexander
  Peysakhovich}.} \bibinfo{year}{2017}\natexlab{}.
\newblock \showarticletitle{Maintaining cooperation in complex social dilemmas
  using deep reinforcement learning}.
\newblock \bibinfo{journal}{\emph{arXiv preprint arXiv:1707.01068}}
  (\bibinfo{year}{2017}).
\newblock


\bibitem[\protect\citeauthoryear{Liebrand and McClintock}{Liebrand and
  McClintock}{1988}]%
        {liebrand1988ring}
\bibfield{author}{\bibinfo{person}{Wim~BG Liebrand} {and}
  \bibinfo{person}{Charles~G McClintock}.} \bibinfo{year}{1988}\natexlab{}.
\newblock \showarticletitle{The ring measure of social values: A computerized
  procedure for assessing individual differences in information processing and
  social value orientation}.
\newblock \bibinfo{journal}{\emph{European journal of personality}}
  \bibinfo{volume}{2}, \bibinfo{number}{3} (\bibinfo{year}{1988}),
  \bibinfo{pages}{217--230}.
\newblock


\bibitem[\protect\citeauthoryear{Littman}{Littman}{1994}]%
        {littman1994markov}
\bibfield{author}{\bibinfo{person}{Michael~L Littman}.}
  \bibinfo{year}{1994}\natexlab{}.
\newblock \showarticletitle{Markov games as a framework for multi-agent
  reinforcement learning}.
\newblock In \bibinfo{booktitle}{\emph{Machine learning proceedings 1994}}.
  \bibinfo{publisher}{Elsevier}, \bibinfo{pages}{157--163}.
\newblock


\bibitem[\protect\citeauthoryear{Lupu, Cui, Hu, and Foerster}{Lupu
  et~al\mbox{.}}{2021}]%
        {lupu2021trajectory}
\bibfield{author}{\bibinfo{person}{Andrei Lupu}, \bibinfo{person}{Brandon Cui},
  \bibinfo{person}{Hengyuan Hu}, {and} \bibinfo{person}{Jakob Foerster}.}
  \bibinfo{year}{2021}\natexlab{}.
\newblock \showarticletitle{Trajectory diversity for zero-shot coordination}.
  In \bibinfo{booktitle}{\emph{International Conference on Machine Learning}}.
  PMLR, \bibinfo{pages}{7204--7213}.
\newblock


\bibitem[\protect\citeauthoryear{McKee, Bai, and Fiske}{McKee
  et~al\mbox{.}}{2022a}]%
        {mckee2022warmth}
\bibfield{author}{\bibinfo{person}{Kevin~R McKee}, \bibinfo{person}{Xuechunzi
  Bai}, {and} \bibinfo{person}{Susan~T Fiske}.}
  \bibinfo{year}{2022}\natexlab{a}.
\newblock \showarticletitle{Warmth and Competence in Human-Agent Cooperation}.
  In \bibinfo{booktitle}{\emph{Proceedings of the 21st International Conference
  on Autonomous Agents and Multiagent Systems}}. \bibinfo{pages}{898--907}.
\newblock


\bibitem[\protect\citeauthoryear{McKee, Gemp, McWilliams,
  Du{\`e}{\~n}ez-Guzm{\'a}n, Hughes, and Leibo}{McKee et~al\mbox{.}}{2020}]%
        {mckee2020social}
\bibfield{author}{\bibinfo{person}{Kevin~R McKee}, \bibinfo{person}{Ian Gemp},
  \bibinfo{person}{Brian McWilliams}, \bibinfo{person}{Edgar~A
  Du{\`e}{\~n}ez-Guzm{\'a}n}, \bibinfo{person}{Edward Hughes}, {and}
  \bibinfo{person}{Joel~Z Leibo}.} \bibinfo{year}{2020}\natexlab{}.
\newblock \showarticletitle{Social Diversity and Social Preferences in
  Mixed-Motive Reinforcement Learning}. In
  \bibinfo{booktitle}{\emph{Proceedings of the 19th International Conference on
  Autonomous Agents and MultiAgent Systems}}. \bibinfo{pages}{869--877}.
\newblock


\bibitem[\protect\citeauthoryear{McKee, Leibo, Beattie, and Everett}{McKee
  et~al\mbox{.}}{2022b}]%
        {mckee2022quantifying}
\bibfield{author}{\bibinfo{person}{Kevin~R McKee}, \bibinfo{person}{Joel~Z
  Leibo}, \bibinfo{person}{Charlie Beattie}, {and} \bibinfo{person}{Richard
  Everett}.} \bibinfo{year}{2022}\natexlab{b}.
\newblock \showarticletitle{Quantifying the effects of environment and
  population diversity in multi-agent reinforcement learning}.
\newblock \bibinfo{journal}{\emph{Autonomous Agents and Multi-Agent Systems}}
  \bibinfo{volume}{36}, \bibinfo{number}{1} (\bibinfo{year}{2022}),
  \bibinfo{pages}{1--16}.
\newblock


\bibitem[\protect\citeauthoryear{Mnih, Badia, Mirza, Graves, Lillicrap, Harley,
  Silver, and Kavukcuoglu}{Mnih et~al\mbox{.}}{2016}]%
        {mnih2016asynchronous}
\bibfield{author}{\bibinfo{person}{Volodymyr Mnih},
  \bibinfo{person}{Adria~Puigdomenech Badia}, \bibinfo{person}{Mehdi Mirza},
  \bibinfo{person}{Alex Graves}, \bibinfo{person}{Timothy Lillicrap},
  \bibinfo{person}{Tim Harley}, \bibinfo{person}{David Silver}, {and}
  \bibinfo{person}{Koray Kavukcuoglu}.} \bibinfo{year}{2016}\natexlab{}.
\newblock \showarticletitle{Asynchronous methods for deep reinforcement
  learning}. In \bibinfo{booktitle}{\emph{International conference on machine
  learning}}. PMLR, \bibinfo{pages}{1928--1937}.
\newblock


\bibitem[\protect\citeauthoryear{Murphy, Ackermann, and Handgraaf}{Murphy
  et~al\mbox{.}}{2011}]%
        {murphy2011measuring}
\bibfield{author}{\bibinfo{person}{Ryan~O Murphy}, \bibinfo{person}{Kurt~A
  Ackermann}, {and} \bibinfo{person}{Michel~JJ Handgraaf}.}
  \bibinfo{year}{2011}\natexlab{}.
\newblock \showarticletitle{Measuring social value orientation}.
\newblock \bibinfo{journal}{\emph{Judgment and Decision making}}
  \bibinfo{volume}{6}, \bibinfo{number}{8} (\bibinfo{year}{2011}),
  \bibinfo{pages}{771--781}.
\newblock


\bibitem[\protect\citeauthoryear{Oord, Li, and Vinyals}{Oord
  et~al\mbox{.}}{2018}]%
        {oord2018representation}
\bibfield{author}{\bibinfo{person}{Aaron van~den Oord}, \bibinfo{person}{Yazhe
  Li}, {and} \bibinfo{person}{Oriol Vinyals}.} \bibinfo{year}{2018}\natexlab{}.
\newblock \showarticletitle{Representation learning with contrastive predictive
  coding}.
\newblock \bibinfo{journal}{\emph{arXiv preprint arXiv:1807.03748}}
  (\bibinfo{year}{2018}).
\newblock


\bibitem[\protect\citeauthoryear{Perez-Nieves, Yang, Slumbers, Mguni, Wen, and
  Wang}{Perez-Nieves et~al\mbox{.}}{2021}]%
        {perez2021modelling}
\bibfield{author}{\bibinfo{person}{Nicolas Perez-Nieves},
  \bibinfo{person}{Yaodong Yang}, \bibinfo{person}{Oliver Slumbers},
  \bibinfo{person}{David~H Mguni}, \bibinfo{person}{Ying Wen}, {and}
  \bibinfo{person}{Jun Wang}.} \bibinfo{year}{2021}\natexlab{}.
\newblock \showarticletitle{Modelling behavioural diversity for learning in
  open-ended games}. In \bibinfo{booktitle}{\emph{International Conference on
  Machine Learning}}. PMLR, \bibinfo{pages}{8514--8524}.
\newblock


\bibitem[\protect\citeauthoryear{Peysakhovich and Lerer}{Peysakhovich and
  Lerer}{2018}]%
        {peysakhovich2018consequentialist}
\bibfield{author}{\bibinfo{person}{Alexander Peysakhovich} {and}
  \bibinfo{person}{Adam Lerer}.} \bibinfo{year}{2018}\natexlab{}.
\newblock \showarticletitle{Consequentialist conditional cooperation in social
  dilemmas with imperfect information (short workshop version)}. In
  \bibinfo{booktitle}{\emph{Workshops at the Thirty-Second AAAI Conference on
  Artificial Intelligence}}.
\newblock


\bibitem[\protect\citeauthoryear{Rapoport}{Rapoport}{1974}]%
        {rapoport1974prisoner}
\bibfield{author}{\bibinfo{person}{Anatol Rapoport}.}
  \bibinfo{year}{1974}\natexlab{}.
\newblock \showarticletitle{Prisoner’s dilemma—recollections and
  observations}.
\newblock In \bibinfo{booktitle}{\emph{Game Theory as a Theory of a Conflict
  Resolution}}. \bibinfo{publisher}{Springer}, \bibinfo{pages}{17--34}.
\newblock


\bibitem[\protect\citeauthoryear{Schwarting, Pierson, Alonso-Mora, Karaman, and
  Rus}{Schwarting et~al\mbox{.}}{2019}]%
        {schwarting2019social}
\bibfield{author}{\bibinfo{person}{Wilko Schwarting}, \bibinfo{person}{Alyssa
  Pierson}, \bibinfo{person}{Javier Alonso-Mora}, \bibinfo{person}{Sertac
  Karaman}, {and} \bibinfo{person}{Daniela Rus}.}
  \bibinfo{year}{2019}\natexlab{}.
\newblock \showarticletitle{Social behavior for autonomous vehicles}.
\newblock \bibinfo{journal}{\emph{Proceedings of the National Academy of
  Sciences}} \bibinfo{volume}{116}, \bibinfo{number}{50}
  (\bibinfo{year}{2019}), \bibinfo{pages}{24972--24978}.
\newblock


\bibitem[\protect\citeauthoryear{Shapley}{Shapley}{1953}]%
        {shapley1953stochastic}
\bibfield{author}{\bibinfo{person}{Lloyd~S Shapley}.}
  \bibinfo{year}{1953}\natexlab{}.
\newblock \showarticletitle{Stochastic games}.
\newblock \bibinfo{journal}{\emph{Proceedings of the national academy of
  sciences}} \bibinfo{volume}{39}, \bibinfo{number}{10} (\bibinfo{year}{1953}),
  \bibinfo{pages}{1095--1100}.
\newblock


\bibitem[\protect\citeauthoryear{Strouse, McKee, Botvinick, Hughes, and
  Everett}{Strouse et~al\mbox{.}}{2021}]%
        {strouse2021collaborating}
\bibfield{author}{\bibinfo{person}{DJ Strouse}, \bibinfo{person}{Kevin McKee},
  \bibinfo{person}{Matt Botvinick}, \bibinfo{person}{Edward Hughes}, {and}
  \bibinfo{person}{Richard Everett}.} \bibinfo{year}{2021}\natexlab{}.
\newblock \showarticletitle{Collaborating with humans without human data}.
\newblock \bibinfo{journal}{\emph{Advances in Neural Information Processing
  Systems}}  \bibinfo{volume}{34} (\bibinfo{year}{2021}),
  \bibinfo{pages}{14502--14515}.
\newblock


\bibitem[\protect\citeauthoryear{Tang, Yu, Chen, Xu, Wang, Fang, Du, Wang, and
  Wu}{Tang et~al\mbox{.}}{2021}]%
        {tang2021discovering}
\bibfield{author}{\bibinfo{person}{Zhenggang Tang}, \bibinfo{person}{Chao Yu},
  \bibinfo{person}{Boyuan Chen}, \bibinfo{person}{Huazhe Xu},
  \bibinfo{person}{Xiaolong Wang}, \bibinfo{person}{Fei Fang},
  \bibinfo{person}{Simon~Shaolei Du}, \bibinfo{person}{Yu Wang}, {and}
  \bibinfo{person}{Yi Wu}.} \bibinfo{year}{2021}\natexlab{}.
\newblock \showarticletitle{Discovering Diverse Multi-Agent Strategic Behavior
  via Reward Randomization}. In \bibinfo{booktitle}{\emph{International
  Conference on Learning Representations}}.
\newblock


\bibitem[\protect\citeauthoryear{Wang, Hughes, Fernando, Czarnecki,
  Du{\'e}{\~n}ez-Guzm{\'a}n, and Leibo}{Wang et~al\mbox{.}}{2018}]%
        {wang2018evolving}
\bibfield{author}{\bibinfo{person}{Jane~X Wang}, \bibinfo{person}{Edward
  Hughes}, \bibinfo{person}{Chrisantha Fernando}, \bibinfo{person}{Wojciech~M
  Czarnecki}, \bibinfo{person}{Edgar~A Du{\'e}{\~n}ez-Guzm{\'a}n}, {and}
  \bibinfo{person}{Joel~Z Leibo}.} \bibinfo{year}{2018}\natexlab{}.
\newblock \showarticletitle{Evolving intrinsic motivations for altruistic
  behavior}.
\newblock \bibinfo{journal}{\emph{arXiv preprint arXiv:1811.05931}}
  (\bibinfo{year}{2018}).
\newblock


\bibitem[\protect\citeauthoryear{Weibull}{Weibull}{1997}]%
        {weibull1997evolutionary}
\bibfield{author}{\bibinfo{person}{J{\"o}rgen~W Weibull}.}
  \bibinfo{year}{1997}\natexlab{}.
\newblock \bibinfo{booktitle}{\emph{Evolutionary game theory}}.
\newblock \bibinfo{publisher}{MIT press}.
\newblock


\bibitem[\protect\citeauthoryear{Zahavy, Schroecker, Behbahani, Baumli,
  Flennerhag, Hou, and Singh}{Zahavy et~al\mbox{.}}{2022}]%
        {zahavy2022discovering}
\bibfield{author}{\bibinfo{person}{Tom Zahavy}, \bibinfo{person}{Yannick
  Schroecker}, \bibinfo{person}{Feryal Behbahani}, \bibinfo{person}{Kate
  Baumli}, \bibinfo{person}{Sebastian Flennerhag}, \bibinfo{person}{Shaobo
  Hou}, {and} \bibinfo{person}{Satinder Singh}.}
  \bibinfo{year}{2022}\natexlab{}.
\newblock \showarticletitle{Discovering policies with domino: Diversity
  optimization maintaining near optimality}.
\newblock \bibinfo{journal}{\emph{arXiv preprint arXiv:2205.13521}}
  (\bibinfo{year}{2022}).
\newblock


\end{thebibliography}
